\documentclass[conference]{IEEEtran}
\IEEEoverridecommandlockouts
\usepackage{cite}
\usepackage{amsmath,amssymb,amsfonts}
\usepackage{algorithmic}
\usepackage{graphicx}
\usepackage{textcomp}
\usepackage{xcolor}
\usepackage{subfigure}
\usepackage{booktabs}
\usepackage{multirow}
\usepackage{indentfirst}
\usepackage{url}
\usepackage{setspace}

\def\BibTeX{{\rm B\kern-.05em{\sc i\kern-.025em b}\kern-.08em
    T\kern-.1667em\lower.7ex\hbox{E}\kern-.125emX}}

\begin{document}
\title{Double-decker: Productive Backscatter Communication Using a Single Commodity Receiver
{\footnotesize \textsuperscript{}}
\thanks{}
}
\author{
\IEEEauthorblockN{Qiwei Wang, Wei Gong}
\IEEEauthorblockA{\textit{University of Science and Technology of China} }
}
\maketitle
\begin{abstract}
Backscatter communication has attracted significant attention for Internet-of-Things applications due to its ultra-low-power consumption.
The state-of-the-art backscatter systems no longer require dedicated carrier generators and leverage ambient signals as carriers.
However, there is an emerging challenge: most prior systems need dual receivers to capture the original and backscattered signals at the same time for tag data demodulation.
This is not conducive to the widespread deployment of backscatter communication.
To address this problem, we present double-decker, a novel backscatter system that only requires a single commercial device for backscatter communication.
The key technology of double-decker is to divide the carrier OFDM symbols into two parts, which are pilot symbols and data symbols.
Pilot symbols can be used as reference signals for tag data demodulation, thus getting rid of the dependence on the dual receiver structure.
We have built an FPGA prototype and conducted extensive experiments.
Empirical results show that when the excitation signal is 802.11g, double-decker achieves a tag data rate of 35.2kbps and a productive data rate of 38kbps, respectively.
The communication range of double-decker is up to 28m in LOS deployment and 24m in NLOS deployment.
\end{abstract}


\section{Introduction}
In the past decade, the Internet-of-Things (IoT) has developed rapidly.
Nowadays, a large number of wireless sensors are deployed around people.
These sensors are used for applications such as localizing, sensing, tracking, etc.
However, most IoT sensors are usually small in size with limited battery capacity, and inconvenient to charge at will.
Conventional wireless communication technology is unbearable for them.
For example, a typical WiFi transceiver consumes 80mW power to send 75 bytes data per second \cite{IEEEexample:Consumption}.
This is infeasible for battery-free devices.
Fortunately, backscatter communication is an emerging technology for IoT applications since it can satisfy the connectivity requirements while having an ultra-low-power consumption.
\begin{figure}[!t]
\centering
\includegraphics[width=0.8\linewidth]{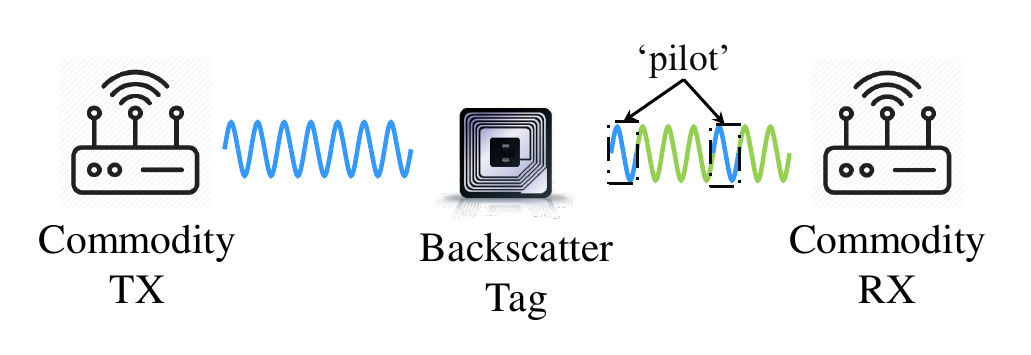}
\caption{Double-decker sets some areas in the carrier that cannot be modulated to retain the necessary information of productive data, which can be used as a reference for tag data decoding.}
\label{idea}
\end{figure}

For a typical wireless communication system, most of the power consumption is caused by the oscillator that generates the carrier signal.
The key idea of backscatter communication is to strip off the extremely energy-intensive high-frequency carrier generator part from battery-free devices.
The excitation signal (i.e., carrier) for backscatter communication is provided by an external carrier generator.
Backscatter tag reflects the excitation signal and modulates its data on the carrier.
The initial backscatter systems usually require dedicated devices to generate excitation signals \cite{kellogg2015wi,iyer2016inter}, these works are excellent but not practical because of the huge hardware overhead.
In recent years, there are many studies to explore leveraging ambient signals for backscatter communication \cite{zhang2016hitchhike,zhang2017freerider,zhao2018spatial,peng2018plora}.
These backscatter systems can uses the ambient signals as carriers to convey tag data.
Therefore, commodity radios can be used as carrier generators and receivers.
Along this line, we can envision the prospect of backscatter communication: leveraging the already-deployed commercial infrastructure and blending with conventional wireless communication technology.
However, although these state-of-the-art backscatter systems are available on commodity radios, as a matter of fact, they are still non-productive communications.
This is because the current backscatter systems cannot recover tag data directly from the backscattered signal.
They have to set an additional receiver to capture the excitation signal for tag data decoding.
Obviously, this structure is incompatible with conventional wireless systems and is not conducive to the widespread deployment of backscatter communication.

To implement a truly productive backscatter system, we have to address a key challenge: {\bfseries How can we demodulate the tag data with a single receiver?}
In this paper, we propose {\bfseries double-decker} with a new modulation mechanism.
As shown in Fig. \ref{idea}, by making concessions in tag data rate, double-decker sets some areas on the carrier as pilot symbols, which are prohibited from data modulation.
By using these pilot symbols as reference signal, double-decker can demodulate the tag data without deploying an additional receiver to capture the excitation signal.
We have built a prototype of double-decker with FPGA and analog RF front-end circuits and comprehensively evaluate the prototype of our design.
The empirical evaluations show that:

\begin{itemize}
\item Different from previous backscatter systems, double-decker only requires a single commodity radio to receive the backscattered signal and can decode tag data and productive data simultaneously. The aggregated throughput of tag data and productive data can reach 73.2 kbps.
\item Double-decker is suitable for multiple protocols such as WiFi, Bluetooth, and ZigBee. The backscatter communication ranges up to 28 m, 22 m, and 20 m for WiFi, Bluetooth, and ZigBee respectively in the LOS scenario, and 24 m, 22 m, and 18 m in NLOS.
\item Double-decker can adjust the rate of tag data and productive data. When the excitation signal is 802.11g, in different modes, we can get the maximal tag data throughput and maximal productive data throughput of 59.5 kbps and 38 kbps respectively.
\end{itemize}
The rest of this paper is organized as follows.
In the next section, we will elaborate on our system design.
In Section III, we will introduce the implementation of our system and evaluate our system comprehensively.
Finally, conclusions and discussions are given in Section IV.

\section{Double-decker Design}
In this section, we will illustrate how we design the double-decker.
To better understand our system, we first take FreeRider as an example to provide a quick review of how the state-of-the-art backscatter system works.
On the basis of FreeRider, we introduce the double-decker.

\subsection{Primer for FreeRider}
FreeRider enables backscatter communication with commodity radios.
The key technique used in FreeRider is {\bfseries codeword translation}.
A codeword is a signal symbol used at the physical layer to represent a special bit sequence \cite{zhang2016hitchhike}.
A codebook is a collection of available codewords.
For a certain modulation method, the codewords of the same codebook are related to each other, that is, they can be converted into each other.
For example, 1Mbps 802.11b WiFi uses binary phase-shift keying (BPSK) modulation which has only two valid codewords, \(code_0\) and \(code_1\), to encode bit zero and bit one respectively.
These two codewords can be expressed as following:
\begin{equation}\nonumber
\begin{split}
codeword_0 &= e^{j\pi\phi_{0}}  \\
codeword_1 &= e^{j\pi\phi_{1}}
\label{codeword}
\end{split}
\end{equation}
Where $\phi_{0}$ represents the phase $0°$ corresponding to $codeword_0$, and $\phi_{1}$ represents the phase $180°$ corresponding to $codeword_1$.
Obviously, there is a $180°$ phase offset between $codeword_0$ and $codeword_1$.
We can change the phase of $codeword_0$ by $180°$ to get $codeword_1$, and vice versa.
Based on this insight, the key idea of codeword translation is to make the backscatter tag act as codeword translator.
By shifting the phase, amplitude, frequency or a combination of them, the backscatter tag can convey its data by transforming a valid codeword to another one.
Following this principle, we are able to leverage productive signals (such as WiFi, Bluetooth, and ZigBee) as carriers for backscatter communication, and the commodity radios can be used as the backscatter transmitters and receivers.
FreeRider also frequency shift the backscattered signals to another channel to avoid the self-interference caused by the excitation signals \cite{iyer2016inter,zhang2016enabling}.

\begin{figure}[!t]
\centering
\subfigure[Architecture of FreeRider. ]{
\centering 
\includegraphics[width=0.45\linewidth]{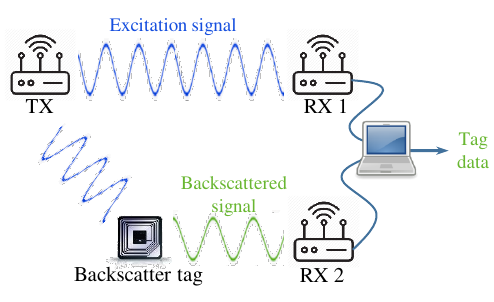}
\label{Previous}
}
\subfigure[Architecture of double-decker. ]{
\centering 
\includegraphics[width=0.45\linewidth]{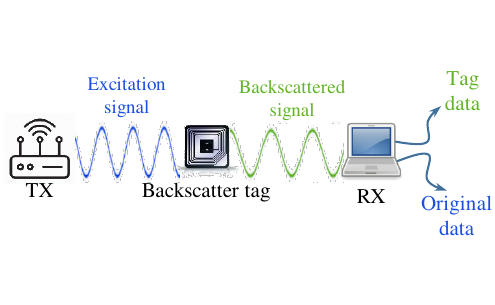}
\label{Double-decker}
}
\caption{Comparison of double-decker and state-of-the-art backscatter systems. }
\label{Architecture}
\end{figure}

Although the codeword translation modulation makes the backscatter systems could leverage ambient signals as carriers, there are still some problems.
It is not difficult to find that the backscattered signal obtained by codeword translation modulation is a combination of the original data and the tag data.
In other words, the receiver cannot directly demodulate the tag data from the backscattered signal.
As Fig. \ref{Previous} shows, in order to recover the bits that the backscatter tag transmits, state-of-the-art backscatter systems need dual receivers to capture excitation signal and backscattered signal respectively.
Then FreeRider can demodulate the tag data by XORing the original data with the backscattered data, which can be written as:
\begin{equation}
tag\ data = original\ data \oplus backscattered\ data
\label{XOR}
\end{equation}

\subsection{Double-Decker}
Obviously, previous backscatter systems allow productive signals as carriers for backscatter communication, but we cannot decode the tag data from backscattered signals independently and the productive data of carriers is also destroyed.
This can not be called real productive communication, so we present double-decker.
The architecture of double-decker is shown in Fig. \ref{Double-decker}.
In this section, we will describe in detail how double-decker works for different excitation signals.

\subsubsection{Backscatter via 802.11b WiFi}
\begin{figure}[!t]
\centering
\includegraphics[width=0.7\linewidth]{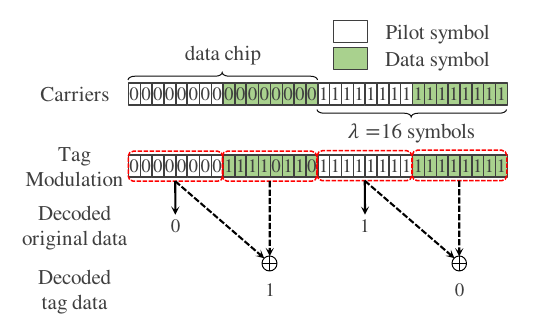}
\caption{Double-decker modulation mechanism: The carrier is divided into many data chips, each data chip is composed of pilot symbols for carrying productive data and data symbols for carrying modulated data. The productive data can be recovered directly from pilot symbols. By XORing the pilot symbols with data symbols, double-decker can decode the tag data.}
\label{modulation_scheme}
\end{figure}

First, we will take 802.11b WiFi as an example to explain the design of double-decker.
Different from prior works, double-decker divides the productive carriers into multiple data chips as Fig. \ref{modulation_scheme} shows.
Each original symbol is spread for \(\lambda\) times to generate the data chip which is a basic unit for embedding tag data, \(\lambda\) is the size of data chip.
The first few symbols are pilot symbols which are prohibited from data modulation.
Then, the rest are data symbols, the backscatter tag is allowed to modulate its data on this part.
For 1Mbps 802.11b signals, each 802.11b symbol has 1 bit of 1 \(\mu s\) long \cite{11b}.
Ideally, the \(\lambda\) can be set to 2 \(\{S_1, S_2\}\), where \(S_1\) is the pilot symbol and \(S_2\) is the data symbol.
However, because of the bursty bit errors during backscatter communication, the decoding performance becomes unstable if the data chip size is short.
As shown in Fig. \ref{modulation_scheme}, ideally, the data symbols should be all 0s or 1s, but in fact, there will be some error bits.
This negative effect causes a demodulating error when the data chip size is short.
So, we use redundancy to solve this problem.
Double-decker empirically set \(\lambda=16\) to achieve a low BER for 802.11b.

Since the data chip (including pilot part and data part) is derived from the same valid symbol, the pilot symbols have the same content as the data symbols.
At the receiver, double-decker can leverage the pilot symbols and data symbols to decode productive data of carrier and tag data respectively.
Because the pilot symbols can not be modulated, receivers can directly recover the productive data of carriers from pilot symbols.
Then, the pilot symbols can serve as the reference signals, that is, the original signals obtained by an additional receiver in FreeRider.
So, double-decker can directly demodulate the tag data by XORing the pilot symbols and data symbols in the same data chip, as shown in (\ref{doubledeckerdecode}).
Thus avoiding dependence on another receiver works on the original channel.

\begin{equation}
tag\ data = pilot\ symbols \oplus data\ symbols
\label{doubledeckerdecode}
\end{equation}

\subsubsection{Backscatter via 802.11g WiFi}
Double-decker's modulation mechanism is also able to support other ambient carriers.
However, for 802.11g signals, the situation is a bit more complicated.
802.11g WiFi uses OFDM modulation, each OFDM symbol contains 24 bits of 4 \(\mu s\) long \cite{11n}.
From the raw data to the OFDM symbol, there are three main processes: scramblering, BCC encoding, and interleaving \cite{11n}.
Scramblering is just a simple bitwise XOR operation, the codeword translation modulation has no effect on it.
The purpose of interleaving is to distribute the bit sequences within an OFDM symbol to discrete subcarriers of the frequency domain.
This process is completely completed within the OFDM symbol and does not interfere with codeword translation modulation.
However, BCC encoding is not so satisfactory.
This process is not completely compatible with codeword translation.
In other words, the decoded data at the receiver are not exact codeword translation's results.

For example, when the \(\lambda\) is set to 4, the data chip of 802.11g has 4 OFDM symbols, of which the first two symbols are used for pilot and the last two are used for data modulation.
Ideally, the pilot symbols should all be 0s, and the data symbols should be all $F$s if it is modulated or all 0s if unmodulated.
In fact, the pilot part of received data contains some ones while the data part has some zeros.
If we still use the XORing result of pilot symbols and data symbols to decode the tag data, this may lead to decoding failure.
Fortunately, we have a critical insight to eliminate this undesirable effect.
We can find that several consecutive bits in the middle part of pilot symbols and data symbols are always the results we expect: the modulated part always $F$s and unmodulated part always 0s.
We empirically set a decoding window of 20 bits length, and use these 20 bits as valid data to present the pilot symbols and data symbols for decoding.
Then, we can correctly decode the productive data and tag data from the backscattered signal.
\begin{figure}[!t]
\centering
\includegraphics[width=\linewidth]{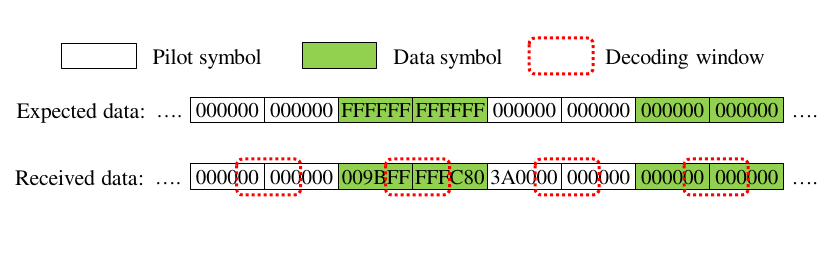}
\caption{Double-decker uses a decoding window to stand for the pilot (or data) symbol as it contains all 0s when unmodulated and all 1s when modulated for decoding.}
\label{OFDMdecoding}
\end{figure}

\subsubsection{Backscatter via ZigBee}
\begin{figure}[!t]
\centering
\includegraphics[width=0.6\linewidth]{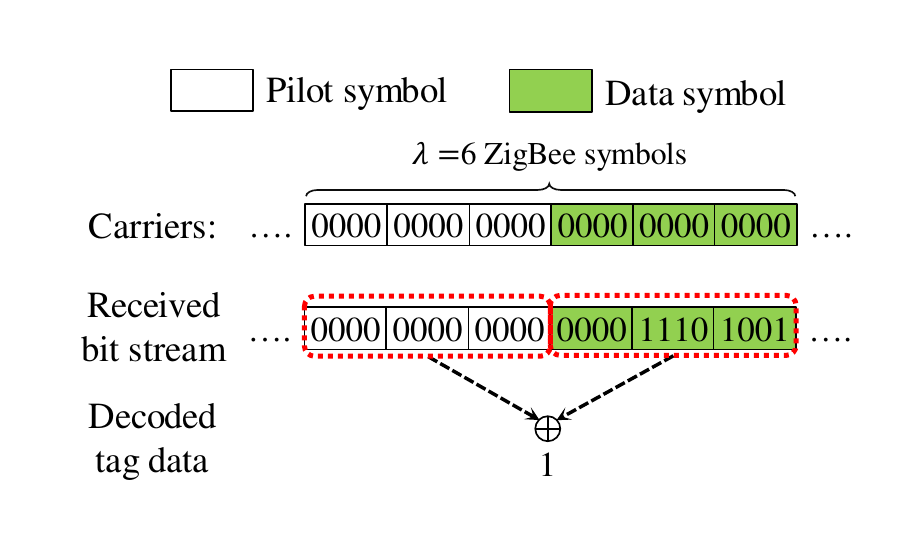}
\caption{The results of modulated ZigBee symbols are not deterministic, but we can still decode the tag data by increasing the size of data chip.}
\label{ZigBeedecoding}
\end{figure}

Each 4-bit ZigBee symbol will be spread to a specified 32-bit long PN sequence predefined in IEEE 802.15.4.
The chip sequences representing each data symbol are modulated onto the carrier using offset quadrature phase-shift keying (O-QPSK) modulation with half-sine pulse shaping \cite{ZigBee}.
Since there is a constant time offset between the in-phase and the quadrature signal, it would damage the half-a-chip offset structure of ZigBee signal if we modulate tag data at bit-level.
So, we modulate tag data based on ZigBee symbols when the excitation signal is ZigBee.
However, the chips obtained by adding $180°$ phase shift to any ZigBee symbol are different from all 16 pre-defined sequences in the protocol.
Usually, ZigBee radios will pick the best-matched sequence to demodulate the phase-shifted symbols so the demodulation results are not deterministic.
As shown in Fig. \ref{ZigBeedecoding}, the first ZigBee symbol in the data part is even not changed.
At this point, when the size of data chip is small, we may not able to decode the tag data.
To solve this problem, we use multiple ZigBee symbols to encode one tag bit to improve the performance.
According to our experiments, double-decker can achieve a BER of \(0.1\%\) when \(\lambda=6\).

\subsubsection{Backscatter via Bluetooth}
Different from the first three excitation signals, BLE uses Gaussian frequency-shift keying (GFSK) modulation.
There are two different frequencies used in BLE, \(f_0\) stands for symbol 0 and \(f_1\) stands for symbol 1 \cite{BLE}.
Typically, for a commodity BLE radio, if the modulation frequency is 1MHz, the frequency interval between \(f_0\) and \(f_1\) is 500kHz.
For BLE codeword translation, it means if we want to modulate tag bit 1, we can produce a frequency shift of 500kHz to turn \(f_0\) to \(f_1\) or vice versa.
If we want to modulate tag bit 0, there is no frequency shift.
Similar to 802.11b, we use redundancy to eliminate the negative effects of bursty errors.
According to our experiments, when \(\lambda=24\), double-decker can achieve a pretty BER performance of \(0.3\%\) for BLE signals.

\subsection{Configuration of Data Chip Mode}
\begin{figure}[!t]
\centering
\includegraphics[width=0.65\linewidth]{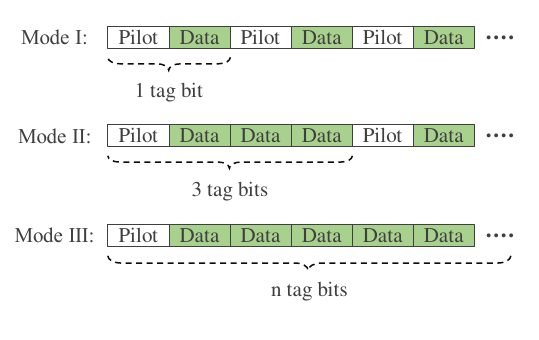}
\caption{Three typical data chip modes with different product data and tag data transmission capabilities.}
\label{pilotmode}
\end{figure}
We just discussed how the double-decker demodulates tag data with a single receiver.
In fact, with the help of pilot symbols, double-decker can also recover the carrier’s productive data from the backscattered signal at the same time.
And double-decker can freely adjust the rate of tag data and carrier's productive data.
Here are three typical data chip setup modes as shown in Fig. \ref{pilotmode}.
In mode 1, each data chip can modulate only one tag data.
At this point, the carrier can be divided into as many data chips as possible.
This mode can maximize the productive rate and it is the default mode in this paper.
We can make each data chip embed more tag data to improve the rate of tag data.
For example, each data chip in mode 2 can embed 3bits of tag data.
So, the tag data rate in mode 2 can reach \(1.5\times\) of that in mode 1, and the productive data rate is only \(50\%\) of that in mode 1.
In extreme cases, as shown in mode 3, except the first symbol which is used as the pilot symbol, the entire payload of the carrier is treated as one data chip.
In this mode, the tag rate is maximized and the double-decker becomes almost the same as FreeRider.

\section{Experimental Evaluation}
We evaluate the end-to-end performance of double-decker in line-of-sight (LOS) and non-line-of-sight (NLOS) deployments.
Above all, we briefly describe the implementation and setup of our experiments:

{\bfseries Transmitters and Receivers:} In our implementation, all transmitters are commodity radios. For WiFi, we use Dell laptops equipped with the Qualcomm Atheros AR938x wireless network adapters as both excitation signal generators and receivers. For Bluetooth, we adopt TI CC2540 commodity radio to send broadcast packets as excitation signals and use TI CC2650 as the receiver. We use a TI CC2530 radio as the transmitter for ZigBee, and the receiver we still use the TI CC2650 which supports both BLE and ZigBee protocols.

{\bfseries Backscatter Tag:} The detector of double-decker tag is constructed using an AD8313 and a TLV3501. The tag modulation circuits and the control circuits are implemented in a XILINX Artix-7 core board. The FPGA uses a mixed-mode clock manager (MMCM) to generate the clock signals with the desired frequency and phase. FPGA uses these clock signals to control the ADG902 to reflect the excitation signal.

\begin{figure}[!t]
\centering
\subfigure[Light-of-sight scenario. ]{
\centering 
\includegraphics[width=0.45\linewidth]{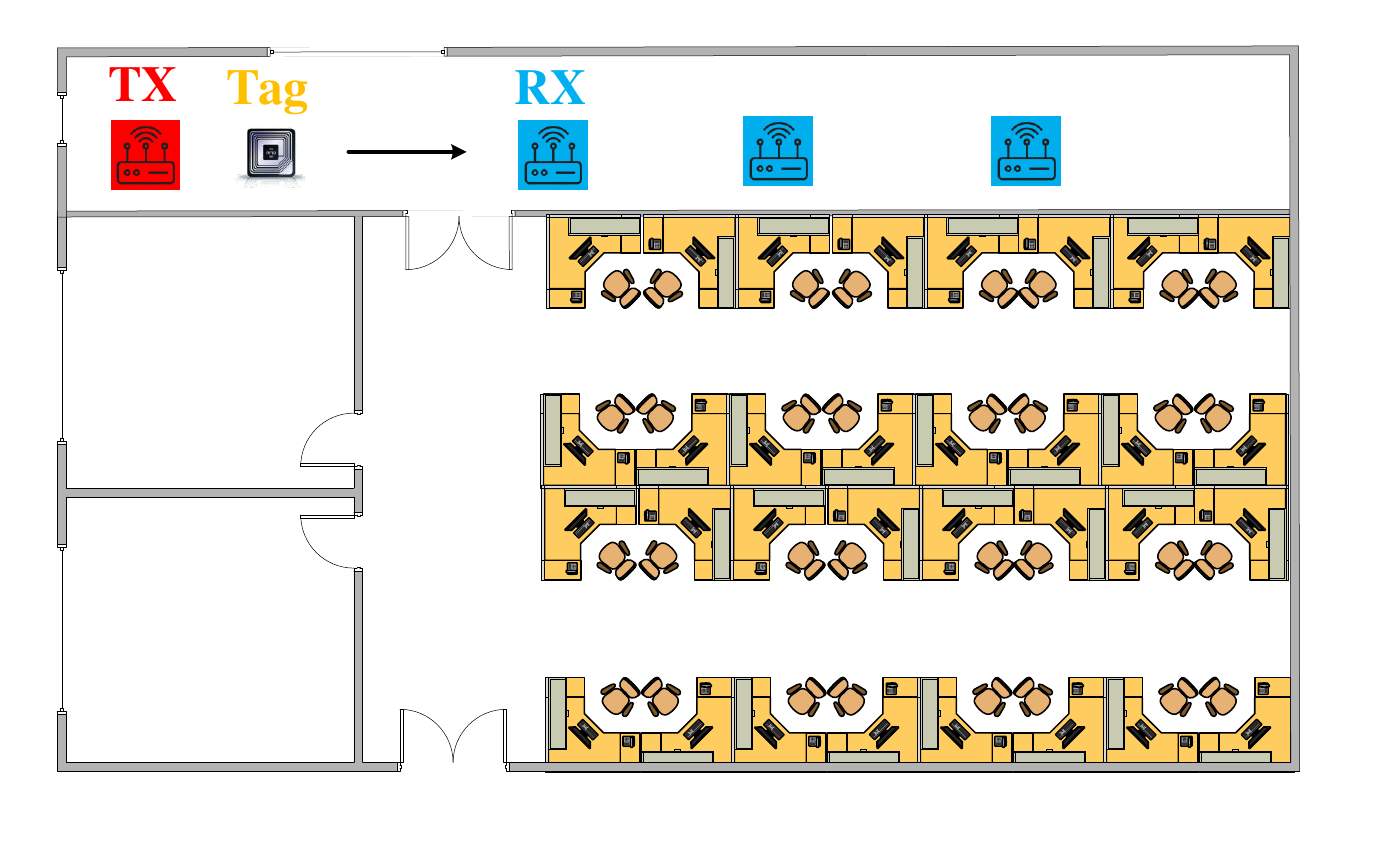}
\label{LOS}
}
\subfigure[Non-light-of-sight scenario. ]{
\centering 
\includegraphics[width=0.45\linewidth]{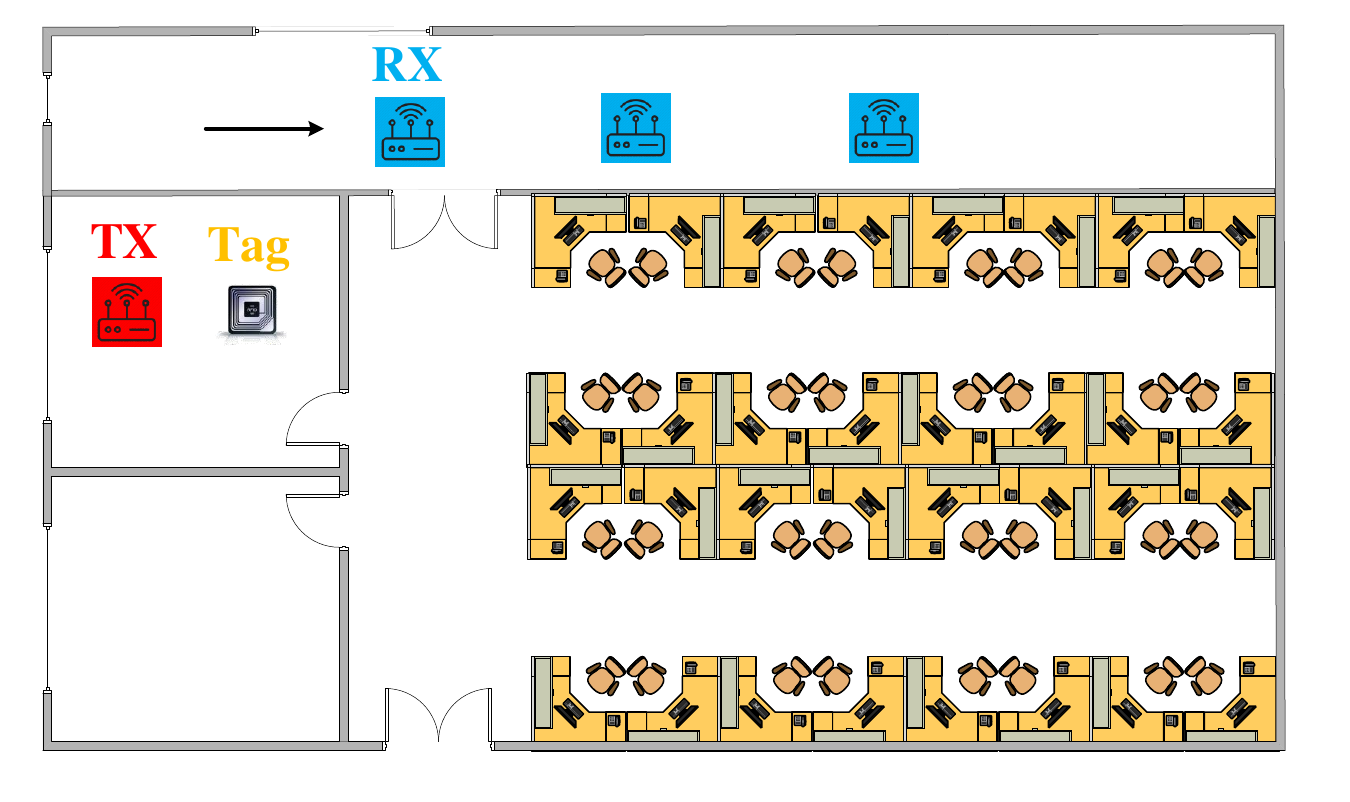}
\label{NLOS}
}
\caption{Experimental setup of our system for LOS and NLOS deployments, the experimental area is 30m*50m. }
\label{Deployments}
\end{figure}

{\bfseries Implementation setup:} Fig. \ref{Deployments} shows the floor plan of our experiments. In LOS deployment, both the transmitter and receivers are placed in the hallway, we fix the distance between double-decker tag and the excitation signal generators to 0.2m. Then we move the receiver gradually away from the tag and measure the received signal strength indicator (RSSI), bit error rate (BER) and throughput of the backscattered signals. In NLOS deployment, we put both the transmitter and double-decker tag in the office room while the receiver is still placed in the hallway.

\begin{figure*}[!t]
\centering
\subfigure[Throughput]{
\centering 
\includegraphics[width=0.3\linewidth]{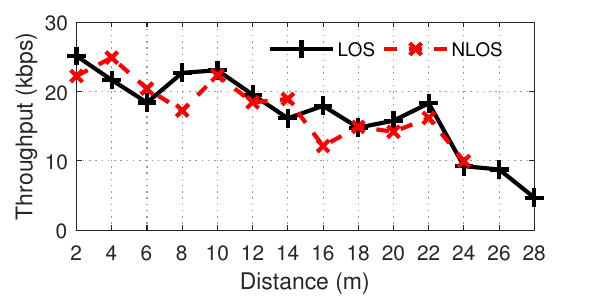}
\label{80211b_TH}
}
\subfigure[BER]{
\centering
\includegraphics[width=0.3\linewidth]{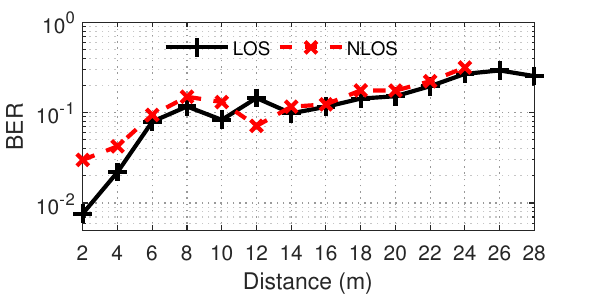}
\label{80211b_BER}

}
\subfigure[RSSI]{
\centering 
\includegraphics[width=0.3\linewidth]{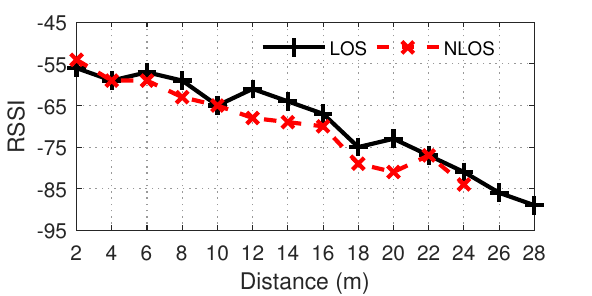}
\label{80211b_RSSI}
}
\caption{Backscatter throughput, BER and RSSI in across distance in LOS and NLOS deployments for WiFi 802.11b.}
\label{80211b}
\end{figure*}
\begin{figure*}[!t]
\centering
\subfigure[Throughput]{
\centering 
\includegraphics[width=0.3\linewidth]{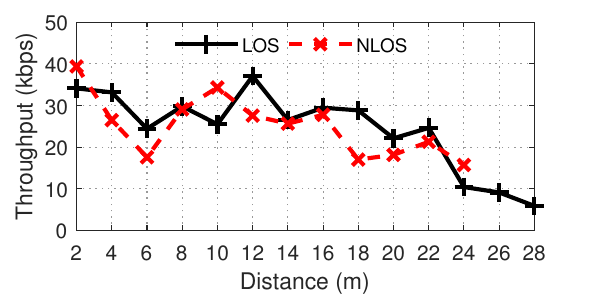}
\label{80211g_TH}
}
\subfigure[BER]{
\centering 
\includegraphics[width=0.3\linewidth]{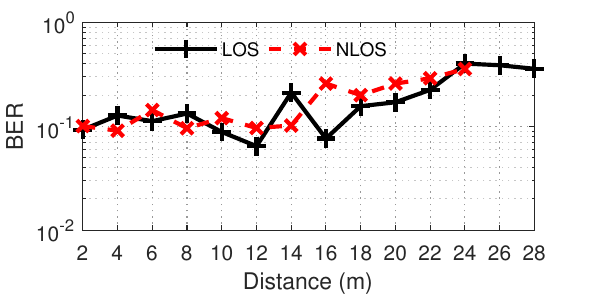}
\label{80211g_BER}
}
\subfigure[RSSI]{
\centering
\includegraphics[width=0.3\linewidth]{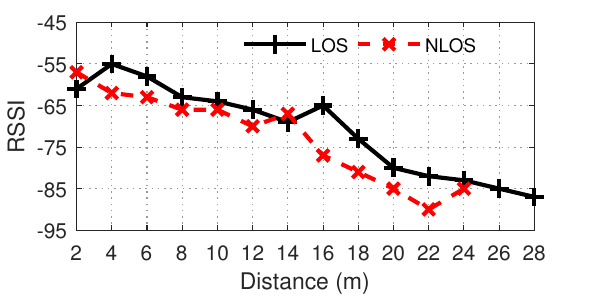}
\label{80211g_RSSI}
}
\caption{Backscatter throughput, BER and RSSI in across distance in LOS and NLOS deployments for WiFi 802.11g.}
\label{80211g}
\end{figure*}

\subsection{Double-decker's Performance}
\subsubsection{Backscatter with 802.11b WiFi}
We first evaluate the end-to-end performance when the excitation signal is 802.11b. We set the transmission rate at 1Mbps for 802.11b and transmit the excitation signal at 15dBm, which is the maximal allowed by our hardware. From Fig. \ref{80211b_TH}, we can see that the maximal backscatter communication ranges of LOS and NLOS deployments are 28 m and 24 m respectively. The maximal throughput of 802.11b can reach 25.1 kbps. Besides, our system achieves above 20 kbps tag data rate when the tag-to-receiver distance is less than 10m. When the receiver moves to a distance over 20 m, the throughput of double-decker is gradually reduced to below 10kbps. However, it can be found that even at the longest distance in NLOS deployment, we can still obtain a throughput of 5 kbps. From Fig. \ref{80211b_RSSI}, we can find that as the distance increases, the RSSI of backscattered signal is attenuated sharply. When the receiver is at 28 m away from the backscatter tag, the strength of the received signal is only -86dBm, which is very close to the commonly considered noise floor of -95dBm. At this point, the receiver can hardly obtain any backscattered packets. In other words, the backscattered signal's received packet rate drops sharply with the increase of distance, and this is the main reason for limiting the distance of backscatter communication.

\subsubsection{Backscatter with 802.11g WiFi}
As Fig. \ref{80211g_TH} shows, when the excitation signal is 802.11g, the maximal distance and RSSI of backscattered signals are exactly the same as 802.11b whether in LOS scenario or NLOS scenario. However, due to the different packet structures, these two signals have obvious differences in throughput. The preamble and header parts of 1Mbps 802.11b packets have a total of 192 bits (i.e., \(192\mu s\)), while the preamble part of 802.11g packets with BPSK modulation only has 5 OFDM symbols (i.e., \(20 \mu s\)). On the other hand, whether the carrier is 802.11b signal or 802.11g signal, the size of data chip is always \(16\mu s\) (16 symbols of 802.11b or 4 OFDM symbols of 802.11g). A shorter preamble means that 802.11g signals can embed more tag data when the packet lengths of the two productive carriers are the same. From Fig. \ref{80211b_TH} and Fig. \ref{80211g_TH}, it is easy to see that the throughput of 802.11g is up to 35.2 kbps, which can be \(1.4\times\) of the tag throughput of 802.11b.

\subsubsection{Backscatter with Bluetooth}
\begin{figure*}[!t]
\centering
\subfigure[Throughput]{
\centering 
\includegraphics[width=0.3\linewidth]{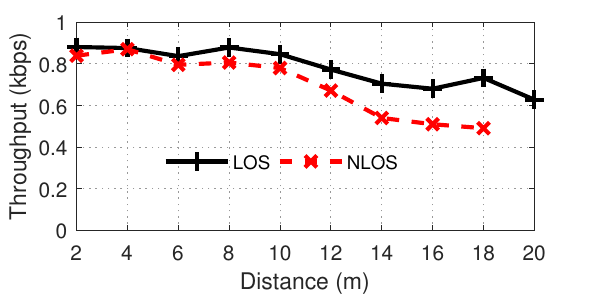}
\label{ble_TH}
}
\subfigure[BER]{
\centering 
\includegraphics[width=0.3\linewidth]{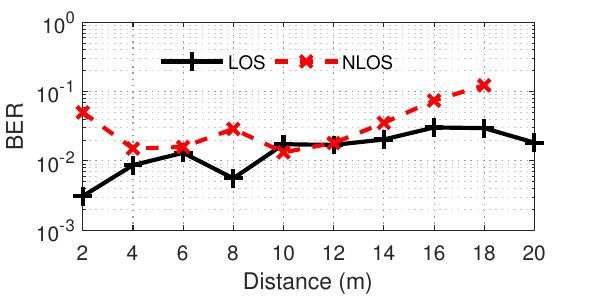}
\label{ble_BER}
}
\subfigure[RSSI]{
\centering 
\includegraphics[width=0.3\linewidth]{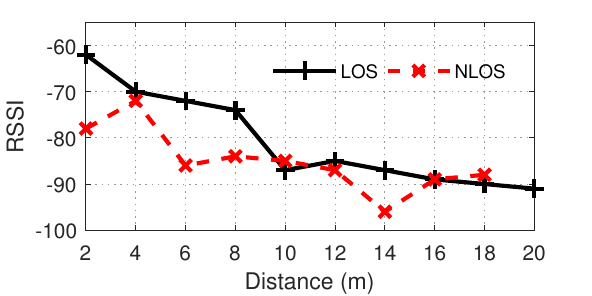}
\label{ble_RSSI}
}
\caption{Backscatter throughput, BER and RSSI in across distance in LOS and NLOS deployments for Bluetooth.}
\label{ble}
\end{figure*}

Then we evaluate the performance of our system when the excitation signal is Bluetooth. Fig. \ref{ble_TH} shows the throughput of double-decker when the receiver moves away from the tag. The receiver can obtain backscattered signals from up to 18 m and 20 m respectively in LOS and NLOS deployments. As shown in Fig. \ref{ble_RSSI}, the received BLE backscattered signal strength degrades quickly to less than -80dBm if the distance exceeds 8 m for LOS deployment. And the RSSI is only -90dBm at 18 m, which is very close to the noise floor. We also achieve the tag data rate of 0.88 kbps, which is a low data rate. According to our analysis, this is due to the limitations of excitation packet rate and packet length. In our experiments, the packet rate of BLE broadcast packet is only 70 packets/second, and the maximal length of the modulatable part of each broadcast packet is only 37 bytes. However, the bit error rate of BLE signals is satisfactory. The BER remains pretty low at a long distance. As shown in Fig. \ref{ble_BER}, even at the farthest distance of backscatter communication for BLE, we can still obtain a BER less than \(3\%\).
\begin{figure*}[!t]
\centering
\subfigure[Throughput]{
\centering 
\includegraphics[width=0.3\linewidth]{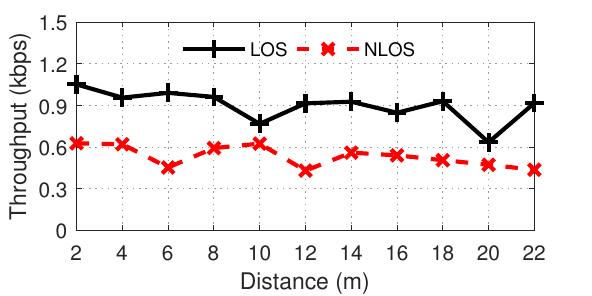}
\label{zigbee_TH}
}
\subfigure[BER]{
\centering 
\includegraphics[width=0.3\linewidth]{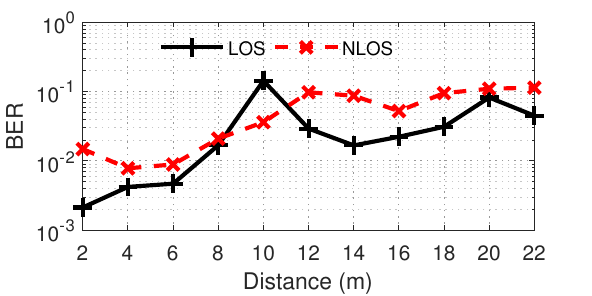}
\label{zigbee_BER}
}
\subfigure[RSSI]{
\centering 
\includegraphics[width=0.3\linewidth]{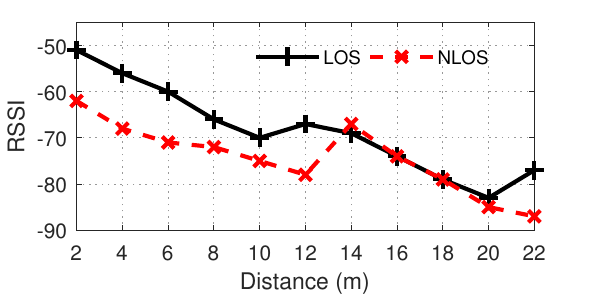}
\label{zigbee_RSSI}
}
\caption{Backscatter throughput, BER and RSSI in across distance in LOS and NLOS deployments for ZigBee.}
\label{zigbee}
\end{figure*}

\subsubsection{Backscatter with ZigBee}
Finally, we have evaluated the performance of our system when the carriers are ZigBee signals. The receiver can decode backscattered packets up to 22 m both in LOS deployment and NLOS deployment. We also have an interesting observation: Among the four protocols, the ZigBee symbol is the longest (\(16\mu s\) for each symbol), while the BER of ZigBee backscattered signal is the lowest. As shown in Fig. \ref{zigbee_BER}, we can achieve \(\sim1e^{-2}\) BER when the distance from the receiver to tag is less than 6 m, and we can still obtain less than \(1\%\) BER when the receiver is 22 m away from the backscatter tag. Besides, from Fig. \ref{zigbee_TH}, we can see that the maximal throughput of ZigBee backscattered signal is 1.05 kbps which is also a low data rate. This is because the packet rate of ZigBee is only 20 packets/second in our experiments. However, this is not a defect of double-decker. Instead, the evaluation results of Bluetooth and ZigBee signals prove that the modulation mechanism of double-decker is universal and can support many ambient signals.
\begin{figure}[!t]
\centering
\includegraphics[width=0.6\linewidth]{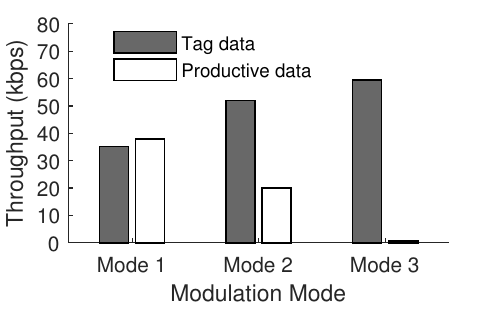}
\caption{Tradeoffs between productive data and tag data throughput under different modes.}
\label{tradeoff}
\end{figure}

\subsection{Tradeoffs of data chip mode}
In Section-II-C, we mentioned that double-decker can adjust the rate of tag data and original data by choosing different data chip modes.
In this part, we take the 802.11g WiFi as an example to evaluate the performance of double-decker in different modes.
The result is shown in Fig. \ref{tradeoff}. We can see that in mode 1, the throughput of tag data and productive data are almost the same.
The aggregated throughput of 802.11g is up to 73.2 kbps, of which the productive data rate and tag data rate are 38 kbps and 35.2 kbps respectively.
In mode 2, the throughput of tag data increases while the throughput of productive data drops significantly.
The tag data rate in mode 2 can reach \(1.478\times\) of that in mode 1, while the productive data rate is only \(52.3\%\) of mode 1, which is consistent with our analysis in Section II.
For mode 3, the throughput of productive data has been degraded to only 0.5 kbps, it is because each packet transmits only one bit of productive data in this mode. At this point, the rate of tag data is up to 59.5 kbps, which is maximal productive data rate of double-decker when excitation signal is 802.11g.

\section{Discussion \& Conclusion}
For the last decade, quite a lot of studies are dedicated to making backscatter communication more versatile, compatible, and easy to deploy.
Along this line and inspired by previous works, we have presented double-decker.
In double-decker, only a single commodity radio is required, we can decode both tag data and productive data simultaneously from the backscattered signal.
According to our experiments, double-decker can achieve a total throughput of tag data and productive data up to 73.2 kbps.
In particular, the productive data rate of carrier can reach 38 kbps in mode three.
It is a significant advancement because we have confirmed that a simple receiver structure can also work in a backscatter system.
In addition, our experiments also illustrate the modulation scheme introduced by double-decker can work with various kinds of ambient carriers, such as WiFi, Bluetooth, and ZigBee.
The design of double-decker reduces the back-to-back requirements of the backscatter system and also makes backscatter tags more compatible with commercial devices.
These improvements made by double-decker can significantly improve the overall performance of the backscatter system and make the backscatter system easier to deploy widely.

\bibliographystyle{IEEEtran}
\bibliography{IEEEexample}

\end{document}